\documentclass[doublecol]{epl2} 

\usepackage{bm,amsmath,amssymb,amsfonts}
\usepackage{graphics}

\title{Tight-binding chains with off-diagonal 
disorder: Bands of extended electronic states induced by 
minimal quasi-one dimensionality}
\shorttitle{Tight-binding chains with off-diagonal 
disorder} 

\author{Atanu Nandy \and Biplab Pal \and Arunava Chakrabarti\thanks{E-mail: \email{arunava\_chakrabarti@yahoo.co.in}}}
\shortauthor{Atanu Nandy \etal}

\institute{Department of Physics, University of Kalyani, Kalyani, West Bengal - 741235, India}

\abstract{It is shown that, an entire class of off-diagonally disordered 
linear lattices composed of two basic 
building blocks and described within a tight binding model
can be tailored to generate {\it absolutely continuous} energy bands. 
It can be achieved 
if linear atomic clusters of an appropriate size are side-coupled 
to a suitable subset of sites in the backbone, and 
if the nearest neighbor hopping integrals, in the 
backbone and in the side-coupled cluster bear a certain ratio. We work out  
the precise relationship between the number of atoms in one of the 
building blocks 
in the backbone, and that in the side-attachment. In addition, we also evaluate 
the definite correlation between the {\it numerical values} of the hopping integrals 
at different subsections of the chain, that can convert an otherwise point spectrum 
(or, a singular continuous one for deterministically disordered lattices) with 
exponentially (or power law) localized eigenfunctions 
to an absolutely continuous 
spectrum comprising one or more bands (subbands) populated by extended, totally transparent 
eigenstates. The results, which are 
analytically exact, put forward a non-trivial variation of the Anderson 
localization [P. W. Anderson, Phys. Rev. \textbf{109}, 1492 (1958)], pointing 
towards its unusual sensitivity to the numerical values of the system parameters 
 and, go well beyond the other related models such as 
the {\it Random Dimer Model} (RDM) [Dunlap \textit{et al.}, Phys. Rev. Lett. \textbf{65}, 88 (1990)].}

\begin{document}

\maketitle

\section{Introduction}
\label{sec1}
Single particle states localize exponentially in a disordered 
system~\cite{anderson,kramer,abrahams}. The effect is strongest in 
one dimension, where there is a complete absence of diffusion 
irrespective of the strength of disorder~\cite{anderson}. 
In two dimensions  
the states retain their exponential decay of amplitude, while in three dimensions 
a possibility of a metal-insulator transition arises. 
The results get adequate support from the  
calculations of the localization length~\cite{rudo1,rudo2}, 
density of states~\cite{alberto} and the multi-fractality of the spectra 
and wave functions of spinless, non-interacting fermionic systems~\cite{rudo3,rudo4,rudo5}. 

The path breaking  
observation by Anderson~\cite{anderson}, over the years, has extended 
its realm well beyond the electronic properties of disordered solid 
materials, and has been found out to be ubiquitous in a wide variety of 
systems. For example, one can refer to the field of localization of light, an idea pioneered 
about three decades ago by Yablonovitch~\cite{yablo} and John~\cite{john}, and 
being carried forward even recently using path-entangled photons~\cite{gilead}
or tailoring of partially coherent light~\cite{svozil}.
Localization of phononic~\cite{montero,vasseur}, polaronic~\cite{barinov,tozer}, 
or plasmonic excitations~\cite{tao,christ,ruting} have also been studied in details and 
have highlighted the general character of Anderson localization induced by disorder. 
From an experimental standpoint, the fundamental issue of localization has been 
substantiated over the past years with the help of   
artificial, tailor made geometries developed by the improved fabrication 
and lithographic techniques. The direct observation of localization of 
matter waves in recent times~\cite{damski,billy,roati} is one such example.

Interestingly,  variations of the canonical case of disorder induced  
Anderson localization have surfaced over the years, particularly, within a tight 
binding description. Resonant tunneling of electronic states in one dimension,  
caused by special short range positional correlation in the so called random dimer 
model (RDM)~\cite{dunlap} initiated such studies. 
Bloch like eigenstates, {\it extended} over the entire lattice were
observed at certain discrete energy eigenvalues rendering the lattice completely 
transparent to an incoming electron possessing such an energy. 
Such a situation was also observed with long range positional correlation in 
one dimension~\cite{moura}, or in quasi-one dimensional ladder networks 
with specially correlated potentials where, the existence of even a 
continuous band of extended states was shown to be possible~\cite{maiti,rudo6}.

In this context, a pertinent question could be, 
is it possible to engineer a complete turnaround, in 
a controlled fashion, in the fundamental character of the energy spectrum 
of non-translationally invariant systems 
such that point-like character of the spectrum, representative of 
localized eigenstates can be converted into an assembly of 
absolutely continuous subbands where only Bloch like eigenstates reside?
The  
existence of continuous bands is reported recently in 
quasi-one dimensional or two dimensional systems with diagonal 
disorder~\cite{maiti,rudo6}. Correlation between the numerical values 
of the hopping integrals in a class of 
topologically disordered quasi-one dimensional closed looped systems has also been 
shown to produce absolutely continuous bands of eigenfunctions recently
~\cite{pal1,pal2}. We thus have a partial answer to the question, 
and it remains to be seen whether such 
bands of extended eigenfunctions can be generated and controlled with 
other variants of an off-diagonally disordered tight binding 
chain of atoms as well.

\begin{figure}[ht]
\centering
\includegraphics[clip,width=8.5cm,angle=0]{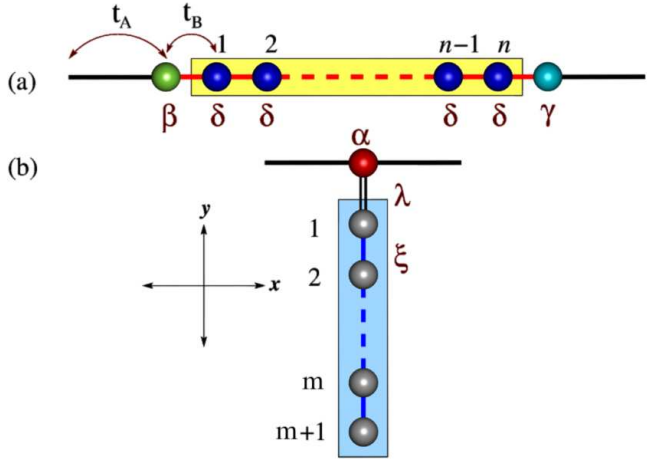}
\caption{(Color online) 
Basic structural units for the off-diagonally disordered chains.
The black and red bonds correspond to the hopping integrals 
$t_A$ and $t_B$. The double bond corresponds to the backbone-side cluster 
coupling $\lambda$ and $\xi$ represents the hopping integral in the 
bulk of the hanging atomic cluster.}
\label{unitcell}
\end{figure}

In this communication we address ourselves this particular question.  
A one dimensional chain is grown along the $x$-direction by placing 
two basic structural units $\beta\delta^n\gamma$ and $\alpha$, with an 
atomic cluster side-coupled to the $\alpha$-sites only (Fig.~\ref{unitcell})
in any desired arrangement, for example, in a completely random way or 
following any deterministically disordered (may be quasiperiodic) 
geometry.

The sites are named in the following way. The $\alpha$ site is flanked by 
two identical bonds $A$. The 
$\beta$ and $\gamma$ sites are flanked by $A$ on the left and $B$ on the right, and 
the other way round. The $\delta$ sites are flanked on either side by the bond of 
type $B$. The hopping integrals corresponding to these bonds are named as 
$t_A$ and $t_B$ respectively. 
The general character of the spectrum in such cases, as can be appreciated 
easily, will reflect localization of electronic states, exponential or 
power law, depending on the distribution.

The side-coupled clusters extend in the $y$-direction and introduces   
a quasi-one dimensionality, but only in the minimal way, and {\it locally} 
at an infinite subset of the atomic sites (of type $\alpha$) in 
the basic chain, henceforth referred to as the `backbone'. We work within 
the framework of the tight binding scheme, and with off-diagonal 
disorder only, that is, the on-site potential is assigned a constant value for 
all the sites, including those in the hanging clusters. 

It is observed, quite contrary to the usual picture of Anderson 
localization~\cite{anderson}, that a definite correlation between the 
{\it numerical values} of the hopping integrals along the backbone ($t_A$ and 
$t_B$),
the backbone-cluster tunnel hopping amplitude ($\lambda$), 
and the intra-cluster hopping ($\xi$) can render any spectrum, namely, point or 
singular continuous in to an assembly of {\it absolutely continuous} subbands.
The continuous subbands turn out to be populated with extended Bloch like states 
only, and this happens {\it irrespective of the electron's energy}, 
in total contrast to the 
already existing results of the RDM class of lattices, where only a finite number of 
resonant eigenstates are observed arising out of a positionally correlated disorder. 
We provide a detailed analysis for a quasiperiodic copper mean lattice~\cite{sil}, 
but emphasize that, the observation is by no means, 
restricted to them. 

Before we conclude this section, 
it may be appropriated to mention at this point that linear 
atomic chains with side-coupled atomic clusters, the so called 
Fano-Anderson defects~\cite{mahan} have drawn interest 
over the past years not only for their unusual localization and 
transport properties, mimicking to some extent, the branched 
polymers~\cite{guinea}, 
but also for their suitability as models of waveguides~\cite{Miroshnichenko}, 
and observation of the Fano resonances 
in the electronic transport~\cite{flach1}. 
Experimental observation of Fano profile in the electronic transmission across 
a quantum wire with a side-coupled quantum dot~\cite{Iye} 
has strengthened the need for 
a detailed study of such systems.
A strong point of interest in such studies 
has been the functionalization of the backbone by the hanging clusters, where 
the electronic states of the side-attachment interfere with the spectrum of the 
linear chain (the backbone) which gives rise to rich spectral features~\cite{grosso1}.
The present work, which allows for a coupling of the discrete eigenstates 
of the hanging cluster with the spectrum of the linear backbone offering a 
spectrum depending on its topography, could be of considerable interest to the 
experimentalists as well.

In what follows, we present the results. In the second and third section 
we provide the tight binding Hamiltonian to work with, and the basic scheme for 
engineering the continuous subbands in the energy spectrum. In the fourth section 
we discuss the case of a copper mean chain (CMC) and illustrates how 
the usual spectrum of a CMC gets converted into a three subband continuous 
pattern as we approach the resonance conditions. In fifth section we present 
the transmission coefficient of finite segments of CMC to corroborate the 
density of states profiles discussed in fourth section, and finally in the 
last section we draw our conclusion.

\begin{figure}[ht]
\centering
\includegraphics[clip,width=8.5cm,angle=0]{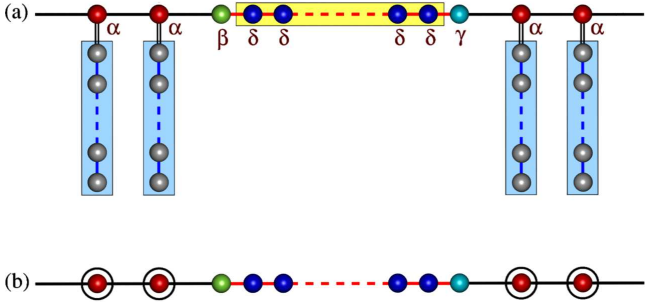}
\caption{(Color online) 
(a) Basic structural units in Fig.~\ref{unitcell} placed at random on the 
backbone. The hopping integrals $t_A$ and $t_B$ are represented by black and 
red lines along the backbone. (b) The hanging sites are decimated to 
generate an effectively one dimensional chain of modified $\alpha$ sites, 
the modification shown by encircling the relevant sites in (a).}
\label{lattice}
\end{figure}

\textbf{The model and the method.} -- 
The variety of {\it atomic environments} is already described in the 
introduction, with reference to Fig.~\ref{unitcell}. 
A typical 
lattice with an arbitrary arrangement of these units 
is illustrated in Fig.~\ref{lattice}(a). 
The array is modeled by the standard tight binding Hamiltonian 
written in the Wannier basis as,
\begin{equation}
H  = \sum_{i} \epsilon_{i} c_{i}^{\dagger} c_{i}
+\sum_{\langle ij \rangle} \left( t_{ij} c_{i}^{\dagger} 
c_{j} 
+ h.c.\right), 
\end{equation}
\label{hamilton}
where the on-site 
potential $\epsilon_i$ at the vertices $\alpha$, $\beta$ and $\gamma$ 
is set equal to a constant, viz., 
$\epsilon_{\alpha}= \epsilon_{\beta}=\epsilon_{\gamma}=\epsilon$. 
The potential at every site of the side 
coupled cluster is designated chosen to be equal to $\epsilon$. The nearest neighbor 
hopping integral $t_{ij} = t_{A}$ or $t_B$ along the backbone, 
and depends on the character of the bond, $A$ or $B$ they 
represent. The tunnel hopping integral between an $\alpha$-site and the first 
site of the side-coupled cluster (shown by a double bond in Fig.~\ref{unitcell}) 
is $t_{ij}=\lambda$, while the intra-cluster hopping in the side-attachment 
is represented by $\xi$.

The fundamental building blocks are placed in any desired pattern on a line 
(the backbone), for example, in a random or a quasiperiodic fashion. 
The geometry can easily be mapped onto a single linear chain comprising 
some effective atoms by decimating the cluster of atomic sites 
in the side-attachment to the $\alpha$-sites. This is illustrated in 
Fig.~\ref{lattice}(b). The decimation process is easily implemented 
through the use of the set of difference equations
\begin{equation}
(E - \epsilon_i) \psi_i = \sum_{\langle ij \rangle} t_{ij} \psi_j,
\label{diff}
\end{equation}
for any $i$-th site in the side-coupled cluster. The resulting linear chain now has 
two kinds of blocks, one being the {\it renormalized} $\alpha$-site having an 
effective energy dependent on-site potential of the form 
\begin{equation}
\tilde{\epsilon}_\alpha  =  \epsilon + \lambda^2 \frac{U_{m}(y)}{\xi U_{m+1}(y)},
\label{renorm}
\end{equation}
and the other is the    
original cluster of $\beta\delta^n\gamma$. In Eq.~\eqref{renorm}
$U_m(y)$ is the $m$-th order Chebyshev polynomial 
of the second kind and $y = (E-\epsilon)/2\xi$. The formulation of 
the equation~\eqref{renorm} is shown in the Appendix in details.

\textbf{Engineering the continuous subbands.} -- 
The effective linear chain depicted in Fig.~\ref{lattice}(b) is described 
by the set of difference equations Eq.~\eqref{diff} with $\epsilon_i = 
\tilde{\epsilon}_\alpha$, $\epsilon_\beta$ or $\epsilon_\gamma$ 
(the latter two values being set equal to $\epsilon$) depending on the 
site. The hopping integral between nearest neighbors $t_{ij}$ 
are still $t_A$ or 
$t_B$ depending on the bond. 
The explicit equations for the three kinds of sites typically look like,
\begin{subequations}
\begin{eqnarray}
(E-\tilde{\epsilon}_\alpha) \psi_{i} & = & t_A \psi_{i+1} + t_A \psi_{i-1},\\ 
(E-\epsilon_\beta) \psi_{i} & = & t_{B} \psi_{i+1} +
t_A \psi_{i-1},\\
(E-\epsilon_\gamma) \psi_{i} & = & t_A \psi_{i+1} +
t_{B} \psi_{i-1},
\label{typicaldiff}
\end{eqnarray}
\end{subequations}
where $i=\alpha$, $\beta$ or $\gamma$, as it comes.
$\tilde{\epsilon}_\alpha$ is given by Eq.~\eqref{renorm}, while 
$\epsilon_\beta = \epsilon_\gamma = \epsilon$ as already explained.

The amplitude of the wave function at any $(i+1)$-th site is related to 
any arbitrary site through a simple product of $2 \times 2$ transfer matrices, and 
is given by,
\begin{equation}
\left (\arraycolsep=0pt \def\arraystretch{1.2} \begin{array}{c}
\psi_{i+1} \\
\psi_{i}  
\end{array} \right )
= \bm{M_i}\!\cdot\! \bm{M_{i-1}}\!\cdots\! \bm{M_2}\!\cdot\! \bm{M_1} 
\left ( \arraycolsep=0pt \def\arraystretch{1.2} \begin{array}{c}
\psi_{1} \\
\psi_{0}  
\end{array} \right).
\end{equation}
The above string of transfer matrices, written explicitly, is a product of 
two basic matrices, viz., $\mbox{\boldmath $M_{\alpha}$}$ and 
$\mbox{\boldmath $M_{\gamma\delta^{n}\beta}$}$
where, 
\begin{subequations}
\begin{align}
&\mbox{\boldmath $M_{\alpha}$} = 
\left( \arraycolsep=5pt \def\arraystretch{1.5} \begin{array}{cccc}
2 x R-\dfrac{\lambda^2 U_m(y)}{\xi t_A U_{m+1}(y)} & -1 \\ 
1 & 0 
\end{array}
\right),\\
&\mbox{\boldmath $M_{\gamma\delta^n\beta}$} =
\left( \arraycolsep=5pt \def\arraystretch{1.5}\begin{array}{cccc}
R U_{n+2}(x) & -\,U_{n+1}(x) \\                  
U_{n+1}(x) & -\,\dfrac{U_{n}(x)}{R}
\end{array}
\right). 
\end{align}
\label{abdgmatrices}
\end{subequations}
Here, $x=(E-\epsilon)/2t_B$, $y=(E-\epsilon)/2\xi$, and $R=t_B/t_A$.
$U_{n}(x)$, as before, represents the $n$-th order Chebyshev polynomial of the 
second kind.
The sequence of the two matrices $\mbox{\boldmath $M_{\alpha}$}$ and
$\mbox{\boldmath $M_{\gamma\delta^{n}\beta}$}$ can be anything, aperiodic or 
\begin{figure}[ht]
\includegraphics[clip,width=8.5cm,angle=0]{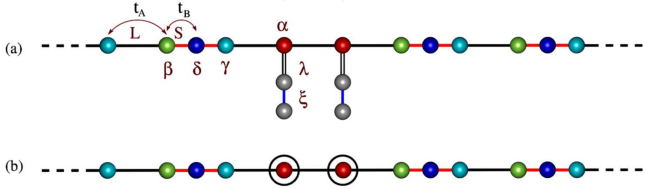}
\caption{(Color online) (a) Portion of an infinite copper mean chain showing 
the four kinds of vertices along with the two bonds as shown. 
(b) Renormalization of the lattice in (a) into a purely 1-d chain. }
\label{cmchain}
\end{figure}
random depending on how the clusters are arranged on the backbone.

It is straightforward to work out the commutator 
\begin{figure}[ht]
\centering
\includegraphics[clip,width=8.5cm,angle=0]{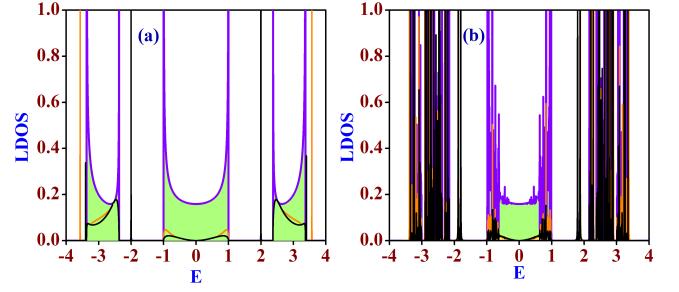}
\caption{(Color online) Spectral landscape of a copper mean chain in (a) \textit{resonance}
condition, (b) $10\%$ deviation from the \textit{resonance} condition. 
The initial values of the parameters in the resonance
condition are
 $\epsilon_{\alpha}=
\epsilon_{\beta}=\epsilon_{\gamma}=\epsilon_{\delta}=0$, $t_A=1$, $t_B=2$, $\lambda=\sqrt{3}$ and 
$\xi=2$. The dangling atoms also have the same on-site potentials as the four kinds of sites of CMC.} 
\label{ldos}
\end{figure}
$[$\mbox{\boldmath $M_{\alpha}$}$,\,$\mbox{\boldmath $M_{\gamma\delta^{n}\beta}$}$]$, and
see that the matrix elements of the commutator read,
$[$\mbox{\boldmath $M_{\alpha}$}$,\,$\mbox{\boldmath $M_{\gamma\delta^{n}\beta}$}$]_{1,1}=0$
and, $[$\mbox{\boldmath $M_{\alpha}$}$,\,$\mbox{\boldmath $M_{\gamma\delta^{n}\beta}$}$]_{2,2} = 0$,
while,
\begin{equation}
\begin{aligned}
\mbox{[\boldmath $M_{\alpha}$},\,\mbox{\boldmath $M_{\gamma\delta^{n}\beta}$}]_{1,2}
  =  R U_{n+2}(x) - 2 x R U_{n+1}(x) + \\
   \frac{U_{n}(x)}{R} + 
\frac{\lambda^2 U_{n+1}(x) U_m(y)}{\xi t_A U_{m+1}(y)},
\label{comm}
\end{aligned}
\end{equation}
with $[$\mbox{\boldmath $M_{\alpha}$}$,\,
$\mbox{\boldmath $M_{\gamma\delta^{n}\beta}$}$]_{2,1}=
[$\mbox{\boldmath $M_{\alpha}$}$,\,$\mbox{\boldmath $M_{\gamma\delta^{n}\beta}$}$]_{1,2}$. 
 
It is interesting to note that if we set $m=n$, that is, if the number of atoms in the 
side-coupled cluster becomes equal to $n+1$, one excess to the number of the 
$\delta$-sites in the cluster $\beta\delta^n\gamma$, the commutator 
$[$\mbox{\boldmath $M_{\alpha}$}$,\,$\mbox{\boldmath $M_{\gamma\delta^{n}\beta}$}$] = 0$.
In terms of the actual lattice it implies that the electronic spectrum will be 
insensitive to the arrangement of the clusters $\alpha$ (renormalized) and $\beta\delta^n\gamma$. 
Any disordered arrangement, deterministic or random, of these two different 
atomic clusters will then, in principle, will be indistinguishable from a perfectly 
periodic arrangement of an infinitely long string of $\alpha$-like sites and 
an infinite array of the $\beta\delta^n\gamma$ polymers.

\textbf{A copper mean quasiperiodic chain as an example.} -- 
To check the arguments laid down so far, 
we fix, without loss of any generality, $n=1$, and 
construct a quasiperiodic copper mean chain (CMC) consisting of 
two `bonds' $A$ and $B$, and 
following the recursive growth rule $A \rightarrow ABB$ and $B \rightarrow A$. 
The resulting CMC has isolated $\alpha$-type atoms flanked by two $A$-bonds on either 
side, and a cluster of $\beta\delta\gamma$ triplets, as shown in Fig.~\ref{cmchain}.
For such a CMC we need to side-couple a two atom cluster to the $\alpha$-sites
(as $m=n$ is the resonance condition).

As we now appreciate the original CMC, under the resonance condition, is equivalent  
to an infinite periodic array of the renormalized $\alpha$ sites (obtained after 
folding the hanging chain back into the backbone site) along with another periodic array 
of the $\beta\delta\gamma$ triplet.
We now evaluate the local density of states (LDOS) 
at the renormalized (encircled) $\alpha$-chain  
and 
any site in the periodic $\beta\delta\gamma$ chain (Fig.~\ref{cmchain}(b)). 
The local densities of states are given, for these two infinite periodic lattices, and 
\begin{figure}[ht]
\centering
\includegraphics[clip,width=8cm,angle=0]{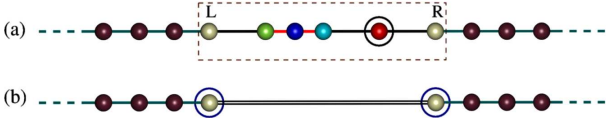}
\caption{(Color online) A finite generation copper mean chain clamped between two 
semi-infinite ordered leads (solid brown circles).} 
\label{translattice}
\end{figure}
for a given set of values of $\epsilon$, $\lambda$, $\xi$, $t_A$ and $t_B$, by
$\rho_{00}^{\alpha} = 1/(\pi \sqrt{Q_\alpha})$ and $\rho_{00}^{\beta} = 
1/(\pi \sqrt{Q_\beta})$ where 
\begin{subequations}
\begin{align}
& Q_\alpha = 4 t_A^2 -
 \left [E - \epsilon - \frac{\lambda^2 U_m(y)}{\xi U_{m+1}(y)} \right ]^2, 
\\
& Q_\beta = \frac{4 t_A^2}{U_{n+1}^2(x)} \nonumber\\
& \quad - t_B^2 \left [ 2 x - \frac{1}{U_n^2(x)} 
\left ( 
U_{n-1}(x) +\frac{R^2 + U_n^2(x)}{R^2 U_{n+1}(x)} \right ) \right ]^2.
\end{align}
\label{cmcdos}
\end{subequations}

As soon as one sets $\lambda = \sqrt{t_B^2 - t_A^2}$, $\xi = t_B$ and $m = n$
the LDOS of the two separate periodic chains 
become identical {\it independent of energy}, resulting in a complete overlap of the 
continuous subbands in the spectra of these individual chains 
under the above resonance condition. 

This is illustrated sequentially in Fig.~\ref{ldos}. 
In the first plot, viz., Fig.~\ref{ldos}(a)
we present the density of states under the resonance condition. The green shaded curves
with violet outlines  
represent the absolutely continuous subbands, and arise out of the $\alpha$-sites. 
The LDOS from the $\beta$, $\delta$ and the $\gamma$ sites cover up the same 
subbands as well. 
The envelops of the black and the orange lines that fall within the green shaded 
zones provide the LDOS at the middle and the end sites in the side-coupled two-atom 
clusters. Thus there is a complete overlap of the absolutely continuous subbands 
arising out of the each individual lattice points. Of course, one should observe the 
appearance of isolated, pinned localized eigenstates, marked by the black and the 
orange spikes in Fig.~\ref{ldos}(a) which occur at
$E=\pm 2$ and $\pm 3.57$. These are the contributions coming from the top and the 
middle atoms in the side-coupled clusters. 

Fig.~\ref{ldos}(b) presents results for the LDOS as we deviate from the resonance 
condition by ten percent. The central continuum practically remains undislodged, while we observe 
dense packing of eigenstates in the subbands at the flanks. We have checked carefully the flow 
of the hopping integrals under the RSRG iterations~\cite{sil} for a wide collection of energies, placed 
densely in all such regions. For every energy belonging to the continuum zone
the hopping integrals keep on oscillating for 
indefinite number of iterations, without 
converging to zero, indicating complete extendedness of the corresponding eigenfunctions.
On the other hand, for $E=\pm 2$ and $\pm 3.57$, the hopping integrals 
$t_A$ and $t_B$ flow to zero under iteration very quickly. As we get non-zero 
densities of states for these energies, the only conclusion that can be drawn 
is that, such states are localized. The small number of iterations indicate 
practically zero overlap of the corresponding wavefunction with the neighboring sites.
This gives us confidence to conclude that such states must be pinned at some 
of the atomic sites in the system, likely places being the hanging clusters themselves.
This has been cross checked by observing the flow of the trace-map, that has been 
used as a diagnostic tool for localization in aperiodic systems~\cite{naumis}. 
For eigenvalues residing 
within the absolutely continuous bands, the trace of the transfer matrix of any 
$l$-th generation CMC remains bounded by $2$~\cite{naumis}, while for the 
localized states it is not.
 This 
observations indicate that a possible experimental growth of such systems can indeed test the 
robustness of the conclusions drawn so far.

\textbf{Two terminal transport.} --
The two terminal transmission coefficient is easily evaluated following the standard
\begin{figure}[ht]
\centering
\includegraphics[clip,width=8.5cm,angle=0]{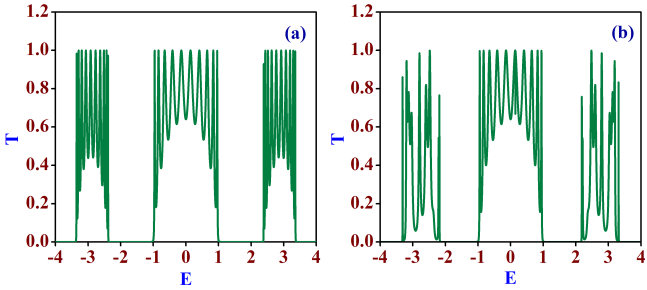}
\caption{(Color online) Transmission spectrum of a $5$-th generation copper mean chain
in (a) \textit{resonance}
condition, (b) $10\%$ deviation from the \textit{resonance} condition.
 The initial values of the parameters in the resonance
condition are
 $\epsilon_{\alpha}=
\epsilon_{\beta}=\epsilon_{\gamma}=\epsilon_{\delta}=0$, $t_A=1$, $t_B=2$, $\lambda=\sqrt{3}$ and 
$\xi=2$. The dangling atoms also have the same on-site potentials as the four kinds of sites of CMC.
The initial values of the lead parameters are respectively $\epsilon_0 = 0$ and $t_0 = 2$.} 
\label{transport}
\end{figure}
prescription~\cite{stone}. A finite generation CMC is clamped between two semi-infinite 
ordered leads (as shown in Fig.~\ref{translattice}) which is characterized 
by a constant on-site potential $\epsilon_0$ and a 
constant nearest neighbor hopping integral $t_0$. The segment of CMC clamped between the 
leads is successively renormalized with the help of the RSRG recursion relations for the 
on-site potentials and the hopping integrals in the CMC exploiting a reversal of its 
growth rule. Without loss of generality, we work out the transmission coefficient for 
odd generation CMC's which `end' with an $A$ bond. To achieve a uniform scaling of the 
end atoms we renormalize the chain by the reverse transformation 
$ABBAA \rightarrow A'$ and $ABB \rightarrow B'$. This makes a $(2n+1)$-th generation 
CMC get folded into the $1$st generation chain (comprising a single $A$-bond) after 
$n$ steps of decimation. The recursion relations 
for the potentials and the hopping integrals are then given by, 
\begin{equation} 
\begin{aligned}
&\tilde{\epsilon}_{\alpha,n+1} = \tilde{\epsilon}_{\alpha,n} + \frac{t_{A,n}^2 \chi_{2,n}}
{\chi_{3,n}} + 
\frac{t_{A,n}^2 \sigma_{2,n}}{\sigma_{3,n}},\\
&\epsilon_{\beta,n+1} = \tilde{\epsilon}_{\alpha,n} +\frac{t_{A,n}^2 \sigma_{2,n}}
{\sigma_{3,n}} + 
\frac{t_{A,n}^2 (E-\epsilon_{\delta,n})}{\sigma_{1,n}},\\
&\epsilon_{\gamma,n+1} = \epsilon_{\gamma,n} + \frac{t_{A,n}^2 \chi_{2,n}}{\chi_{3,n}} + 
\frac{t_{B,n}^2 (E-\epsilon_{\beta,n})}{\sigma_{1,n}},\\
&\epsilon_{\delta,n} = \epsilon_{\gamma,n} + \frac{t_{B,n}^2 (E-\epsilon_{\beta,n})}
{\sigma_{1,n}} + 
\frac{t_{A,n}^2 (E-\epsilon_{\delta,n})}{\sigma_{1,n}},\\
&t_{A,n+1} = \frac{t_{A,n}^3 t_{B,n}^2}{\chi_{3,n}},\\
&t_{B,n+1} = \frac{t_{B,n}^2 t_{A,n}}{\sigma_{1,n}}.
\end{aligned}
\label{recursion1}
\end{equation}
The `end' sites $L$ and $R$ get renormalized following the rules,
\begin{equation}
\begin{aligned}
&\epsilon_{L,n+1} = \epsilon_{L,n} + \frac{t_{A,n}^2 \chi_{2,n}}{\chi_{3,n}},\\
&\epsilon_{R,n+1} = \epsilon_{R,n} + \frac{t_{A,n}^2 \sigma_{2,n}}{\sigma_{3,n}}. 
\end{aligned}
\label{endrecur}
\end{equation}
Here, $\chi_{2,n} = [(E-\epsilon_{\delta_n})\chi_{1,n} - t_{B,n}^2 (E-\epsilon_{\alpha,n})]$, with $\chi_{1,n}=(E-\epsilon_{\gamma,n}) (E-\epsilon_{\alpha,n}) - t_{A,n}^2$.
$\chi_{3,n} = [(E-\epsilon_{\beta,n}) \chi_{2,n} - t_{B,n}^2 \chi_{1,n}]$, 
$\sigma_{1,n}=(E-\epsilon_{\delta,n}) (E-\epsilon_{\beta,n}) - t_{B,n}^2$, 
$\sigma_{2,n}=(E-\epsilon_{\gamma,n}) \sigma_{1,n} - t_{B,n}^2 (E-\epsilon_{\beta,n})$, 
and, $\sigma_{3,n} = \sigma_{2,n} (E-\epsilon_{\alpha,n}) - \sigma_{1,n} t_{A,n}^2$.

The two terminal 
transport coefficient of a $(2n+1)$-th generation CMC is then given by, 
\begin{eqnarray}
&T=\dfrac{4\sin^{2}ka}{|\mathcal{P}|^{2}+|\mathcal{Q}|^{2}}. \\
&\text{with}\quad \mathcal{P}=[(M_{12}-M_{21})+(M_{11}-M_{22})\cos ka] \nonumber\\
&\text{and}\quad \mathcal{Q}=[(M_{11}+M_{22})\sin ka],\nonumber
\label{transformula}
\end{eqnarray}
where $M_{11}=(E-\epsilon_{R,n})(E-\epsilon_{L,n})/t_0t_{A,n}-t_{A,n}/t_0$, 
$M_{12}=-(E-\epsilon_{R,n})/t_{A,n}$, $M_{21}=(E-\epsilon_{L,n})/t_{A,n}$, and 
$M_{22}=-t_0/t_{A,n}$.

In Fig.~\ref{transport} we exhibit the two terminal transport of a $5$-th generation CMC with the $\alpha$-sites attached with two-atom clusters. Three clear transmission windows are visible 
exactly over the energy ranges of the absolutely continuous subbands when resonance condition is 
enforced (Fig.~\ref{transport}(a)). This is at per with the expectation that these subbands are 
populated with extended eigenfunctions only. Interestingly, even with a ten percent deviation 
from the condition of resonance the transparent windows of transmission coefficient demonstrates 
the robustness of the result.

It is of interest to observe that, the renormalized potential 
$\tilde{\epsilon}_\alpha$ reduces to its bare scale value $\epsilon_\alpha = 
\epsilon$ for energy eigenvalues which are solutions of the equation 
$U_m(y) = 0$. 
Each such energy values, which reside within the energy band 
of the semi-infinite leads, will give rise to a resonant electronic transmission.
For these energies, the excursion of the propagating electron in the 
hanging clusters does not change the phase of the wavefunction. Similar 
issues, including the occurrence of the stop bands in between the absolute 
continua of the spectrum of a model quantum wire with side-coupled 
nanowires were discussed previously by Orellana \textit{et al.}~\cite{orellana}

It is also important to appreciate that, the resonant transmission ($T=1$) 
seen in the present case is definitely caused by the fact that the transfer matrices 
across the clusters $\alpha$ and $\beta\delta^n\gamma$ commute irrespective of 
energy when we set the desired correlation between the hopping integrals, as 
mentioned already. This is to be contrasted with the case of, for example, the 
RDM, where under a resonance condition, the transfer matrix across  a local 
pair of impurity atoms just turned out to be an identity matrix~\cite{dunlap}
at only a special value of the electron's energy. In the present case, the commutation 
makes the lattice indistinguishable from a periodic array of the same scatterers. 
The transmission resonances thus result out of the phase coherence, as the 
electron is made to travel through the system.

Before we end this section, it is pertinent to remind the reader that the choice of a copper mean 
lattice is just for the sake of presenting analytically exact results. The condition of resonance, and 
it's consequences, are by no means restricted to such a specific case, and definitely hold 
good for a completely random distribution of the clusters $\beta\delta^n\gamma$ and $\alpha$. Variants 
of the idea presented in this communication has already been tested with other geometries elsewhere~\cite{pal1}.

\textbf{Concluding remarks.} --
We have shown that suitably introducing minimal quasi-one dimensionality 
to a selected subset of atomic sites in an infinite linear chain one can obtain 
absolutely continuous spectrum in an off-diagonal model with random or any kind 
of deterministic disorder. This requires an appropriate correlation between the 
numerical values of a subset of the Hamiltonian parameters, in contradistinction 
with the conventional Anderson localization problem. The results indicate a possibility 
of manipulating a spectral crossover, if not an insulator to metal transition by tuning 
the hopping integrals appropriately. The result appears to be robust even when one deviates 
from the ideal conditions of resonance, a fact that may inspire experimentalists to undertake 
experiments with quantum wires, for example, with quantum dots coupled to it from a side.

\acknowledgments
A.N. is indebted to UGC, India for the financial support provided through a research fellowship 
[Award letter no. F.$17$-$81 / 2008$ (SA-I)]. 
B.P. is grateful to DST, India for an INSPIRE Fellowship. A.C. acknowledges 
financial support from DST, India through a PURSE grant.

\textbf{Appendix.} --
\setcounter{equation}{0}
\def\theequation{A.$\,$\arabic{equation}}
For derivation of the Eq.~\eqref{renorm}, let us first look at the Fig.~\ref{unitcell}(b).
From this figure, we can have,
\begin{equation}
\left (\arraycolsep=0pt \def\arraystretch{1.2} \begin{array}{c}
\psi_{m+1} \\
\psi_{m}  
\end{array} \right )
=
\bm{M_m}\!\cdot\! \bm{M_{m-1}}\!\cdots\! \bm{M_3}\!\cdot\! \bm{M_2} 
\left ( \arraycolsep=0pt \def\arraystretch{1.2} \begin{array}{c}
\psi_{2} \\
\psi_{1}  
\end{array} \right).
\end{equation}
From which we must have,
\begin{equation}
\left (\arraycolsep=0pt \def\arraystretch{1.2} \begin{array}{c}
\psi_{m+1} \\
\psi_{m}  
\end{array} \right )
=
\bm{M^{m-1}}
\left ( \arraycolsep=0pt \def\arraystretch{1.2} \begin{array}{c}
\psi_{2} \\
\psi_{1}  
\end{array} \right).
\end{equation}
Thus, from the above equation, writing in terms of the Chebyshev polynomial we obtain,
\begin{equation}
\left (\arraycolsep=0pt \def\arraystretch{1.2} \begin{array}{c}
\psi_{m+1} \\
\psi_{m}  
\end{array} \right )
= 
\left( \arraycolsep=5pt \def\arraystretch{1.5} \begin{array}{cccc}
2 y U_{m-2}(y) - U_{m-3}(y) & -U_{m-2}(y) \\ 
U_{m-2}(y) & -U_{m-3}(y) 
\end{array}
\right)
\left ( \arraycolsep=0pt \def\arraystretch{1.2} \begin{array}{c}
\psi_{2} \\
\psi_{1}  
\end{array} \right).
\end{equation}
This obviously gives,
\begin{equation}
\left (\arraycolsep=0pt \def\arraystretch{1.2} \begin{array}{c}
\psi_{m+1} \\
\psi_{m}  
\end{array} \right )
=
\left( \arraycolsep=5pt \def\arraystretch{1.5} \begin{array}{cccc}
U_{m-1}(y) & -U_{m-2}(y) \\ 
U_{m-2}(y) & -U_{m-3}(y) 
\end{array}
\right)
\left ( \arraycolsep=0pt \def\arraystretch{1.2} \begin{array}{c}
\psi_{2} \\
\psi_{1}  
\end{array} \right).
\label{appendixeqn1}
\end{equation}
Here, $y = (E-\epsilon)/2 \xi$.
From Eq.~\eqref{appendixeqn1}, we can write,
\begin{subequations}
\begin{eqnarray}
\psi_{m+1} & = & U_{m-1}(y) \psi_{2} - U_{m-2}(y) \psi_{1},\\ 
\psi_{m} & = & U_{m-2}(y) \psi_{2} - U_{m-3}(y) \psi_{1}.
\end{eqnarray}
\label{appendixeqn2}
\end{subequations}
Now, if we write down the difference equation for the $(m+1)$-th atom in the hanging cluster, we will get,
\begin{equation}
(E-\epsilon) \psi_{m+1} = \xi \psi_{m}.
\label{appendixeqn3}
\end{equation}
Therefore, substituting Eq.~\eqref{appendixeqn2} into the Eq.~\eqref{appendixeqn3} one
can easily obtain after simplification the following expression.
\begin{equation}
\psi_{2} = \dfrac{U_{m-1}(y)}{U_{m}(y)} \psi_{1}.
\label{appendixeqn4}
\end{equation}
Now the difference equation for the $1$st atomic site in the hanging cluster reads as,
\begin{equation}
(E-\epsilon) \psi_{1} = \lambda \psi_{\alpha} + \xi \psi_{2}.
\label{appendixeqn5}
\end{equation}
By the use of Eq.~\eqref{appendixeqn4}, one can obtain from Eq.~\eqref{appendixeqn5},
\begin{equation}
\psi_{1} = \dfrac{\lambda U_{m}(y)}{\xi U_{m+1}(y)} \psi_{\alpha}.
\label{appendixeqn6}
\end{equation}
Finally the difference equation for the $\alpha$-kind of site is given by,
\begin{equation}
(E-\epsilon) \psi_{\alpha} = \lambda \psi_{1} + t_A \sum_{j} \psi_j.
\label{appendixeqn7}
\end{equation}
From Eq.~\eqref{appendixeqn6} and Eq.~\eqref{appendixeqn7}, we get the final form of the
renormalized on-site potential as, 
\begin{equation}
\tilde{\epsilon}_{\alpha} = \epsilon + \lambda^2 \frac{U_{m}(y)}{\xi U_{m+1}(y)}.
\end{equation}

%


\begin{thebibliography}{99}

\bibitem{anderson}
  \Name{Anderson P. W.}
  \REVIEW{Phys. Rev.}{109}{1958}{1492}.
  
\bibitem{kramer}
  \Name{Kramer B. \and MacKinnon A.}
  \REVIEW{Rep. Prog. Phys.}{56}{1993}{1469}.

\bibitem{abrahams}
  \Name{Abrahams E., Anderson P. W., Licciardello D. C. \and Ramakrishnan T. V.}
  \REVIEW{Phys. Rev. Lett.}{42}{1979}{673}.
  
\bibitem{rudo1}
  \Name{R\"{o}mer R. A. \and Schulz-Baldes H.}
  \REVIEW{Europhys. Lett.}{68}{2004}{247}.
  
\bibitem{rudo2}
  \Name{Eilmes A., R\"{o}mer R. A. \and Schreiber M.}
  \REVIEW{Physica B}{296}{2001}{46}.  
  
\bibitem{alberto}
  \Name{Rodr\'{i}guez A.}
  \REVIEW{J. Phys. A: Math. Gen.}{39}{2006}{14303}. 
  
\bibitem{rudo3}
  \Name{Rodriguez A., Vasquez L. J., \and R\"{o}mer R. A.}
  \REVIEW{Phys. Rev. B}{78}{2008}{195107}.  

\bibitem{rudo4}
  \Name{Rodriguez A., Vasquez L. J., Slevin K. \and R\"{o}mer R. A.}
  \REVIEW{Phys. Rev. B}{84}{2011}{134209}.
  
\bibitem{rudo5}
  \Name{Pinski S. D., Schirmacher W. \and R\"{o}mer R. A.}
  \REVIEW{Europhys. Lett.}{97}{2012}{16007}.  
  
\bibitem{yablo}
  \Name{Yablonovitch E.}
  \REVIEW{Phys. Rev. Lett.}{58}{1987}{2059}.  
  
\bibitem{john}
  \Name{John S.}
  \REVIEW{Phys. Rev. Lett.}{58}{1987}{2486}.
  
\bibitem{gilead}
  \Name{Gilead Y., Verbin M. \and Silberberg Y.}
  \REVIEW{Phys. Rev. Lett.}{115}{2015}{133602}.
  
\bibitem{svozil}
  \Name{Svozil\'{i}k J., Pe\v{r}ina J., Jr., Slevin K. \and Torres J. P.}
  \REVIEW{Phys. Rev. A}{89}{2014}{053808}.
  
\bibitem{montero}
  \Name{Montero de Espinosa F. R., Jim\'{e}nez E. \and Torres M.}
  \REVIEW{Phys. Rev. Lett.}{80}{1998}{1208}.  
  
\bibitem{vasseur}
  \Name{Vasseur J. O., Deymier P. A., Frantziskonis G., 
  Hong G., Djafari-Rouhani B. \and  Dobrzynski L.}
  \REVIEW{J. Phys.: Condens. Matter}{10}{1998}{6051}.
  
\bibitem{barinov}  
   \Name{Barinov I. O., Alodjants A. P. \and Arakelian S. M.}
  \REVIEW{Quantum Electron.}{39}{2009}{685}.  
  
\bibitem{tozer}
  \Name{Tozer O. R. \and Barford W.}
  \REVIEW{Phys. Rev. B}{89}{2014}{155434}. 

\bibitem{tao}
  \Name{Tao A., Sinsermsuksaul P. \and Yang P.}
  \REVIEW{Nat. Nanotechnol.}{2}{2007}{435}.

\bibitem{christ}
  \Name{Christ A., Ekinci Y., Solak H. H., Gippius N. A., 
  Tikhodeev S. G. \and Martin O. J. F.}
  \REVIEW{Phys. Rev. B}{76}{2007}{201405(R)}. 
  
\bibitem{ruting}
  \Name{R\"{u}ting F.}
  \REVIEW{Phys. Rev. B}{83}{2011}{115447}.
    
\bibitem{damski}
  \Name{Damski B., Zakrzewski J., Santos L., Zoller P. \and Lewenstein M.}
  \REVIEW{Phys. Rev. Lett.}{91}{2003}{080403}.    

\bibitem{billy}
  \Name{Billy J., Josse V., Zuo Z., Bernard A., Hambrecht B., 
  Lugan P., Cl\'{e}ment D., Sanchez-Palencia L., Bouyer P. \and Aspect A.}
  \REVIEW{Nature (London)}{453}{2008}{891}.
  
\bibitem{roati}
  \Name{Roati G., D'Errico C., Fallani L., Fattori M., Fort C., 
  Zaccanti M., Modugno G., Modugno M. \and Inguscio M.}
  \REVIEW{Nature (London)}{453}{2008}{895}.
  
\bibitem{dunlap}
  \Name{Dunlap D. H., Wu H-L. \and Phillips P. W.}
  \REVIEW{Phys. Rev. Lett.}{65}{1990}{88}.
  
\bibitem{moura}
  \Name{de Moura F. A. B. F. \and Lyra M. L.}
  \REVIEW{Phys. Rev. Lett.}{81}{1998}{3735}.  

\bibitem{maiti}
  \Name{Sil S., Maiti S. K. \and Chakrabarti A.}
  \REVIEW{Phys. Rev. B}{78}{2008}{113103}.

\bibitem{rudo6}
  \Name{Rodriguez A., Chakrabarti A. \and R\"{o}mer R. A.}
  \REVIEW{Phys. Rev. B}{86}{2012}{085119}.
  
\bibitem{pal1}
  \Name{Pal B., Maiti S. K. \and Chakrabarti A.}
  \REVIEW{Europhys. Lett.}{102}{2013}{17004}.

\bibitem{pal2}
  \Name{Pal B. \and Chakrabarti A.}
  \REVIEW{Phys. Lett. A}{378}{2014}{2782}.
  
\bibitem{sil}
  \Name{Sil S., Karmakar S. N., Moitra R. K. \and Chakrabarti A.}
  \REVIEW{Phys. Rev. B}{48}{1993}{4192(R)}.
%
\bibitem{mahan}
	\Name{Mahan G. D.}
	\Book{Many-Particle Physics}
	\Publ{Plenum Press, New York}
	\Year{1993}.
%
\bibitem{guinea}
	\Name{Guinea F. \and Verg\'{e}s J. A.}
	\REVIEW{Phys. Rev. B}{35}{1987}{979}.
%
\bibitem{Miroshnichenko}
	\Name{Miroshnichenko A. E. \and Kivshar Y. S.}
	\REVIEW{Phys. Rev. E}{72}{2005}{056611}.
%
\bibitem{flach1}
	\Name{Miroshnichenko A. E., Flach S. \and Kivshar Y. S.}
	\REVIEW{Rev. Mod. Phys.}{82}{2010}{2257}.
%
\bibitem{Iye}
	\Name{Kobayashi K., Aikawa H., Sano A., Katsumoto S. \and Iye Y.}
	\REVIEW{Phys. Rev. B}{70}{2004}{035319}.
%
\bibitem{grosso1}
	\Name{Farchioni R., Grosso G. \and Parravicini G. P.}
	\REVIEW{Phys. Rev. B}{85}{2012}{165115}.
%
\bibitem{naumis}
	\Name{Naumis G. G.}
	\REVIEW{J. Phys.: Condens. Matter}{15}{2003}{5969}.

\bibitem{stone}
  \Name{Stone A. D., Joannopoulos J. D. \and Chadi D. J.}
  \REVIEW{Phys. Rev. B}{24}{1981}{5583}.
%
\bibitem{orellana}
	\Name{Orellana P. A., Guevara M. L. Ladron de, Dominguez-Adame F. \and Gomez I.}
	\REVIEW{Phys. Status Solidi C}{1}{2004}{S50}.  
\end{thebibliography}
\end{document}